\documentclass[english,5p,twocolumn,times]{elsarticle}
\usepackage[T1]{fontenc}
\usepackage[utf8]{inputenc}
\usepackage{geometry}
\usepackage{babel}
\usepackage{array}
\usepackage{verbatim}
\usepackage{textcomp}
\usepackage{amstext}
\usepackage{graphicx}
\usepackage[unicode=true]
 {hyperref}

\makeatletter

\providecommand{\tabularnewline}{\\}

\newenvironment{lyxlist}[1]
	{\begin{list}{}
		{\settowidth{\labelwidth}{#1}
		 \setlength{\leftmargin}{\labelwidth}
		 \addtolength{\leftmargin}{\labelsep}
		 }}
	{\end{list}}

\usepackage{color}
\usepackage{graphics}
\usepackage{array}
\usepackage{graphicx}
\usepackage{lastpage}
\usepackage{float}

\usepackage{times}

\usepackage{lineno}

\journal{XXX}

\usepackage{amsmath,amssymb}

\usepackage{colortbl}
\usepackage{multicol}

\definecolor{v1}{rgb}{0.0,0.8,0.0}
\definecolor{v2}{rgb}{0.8,1.0,0.0}
\definecolor{v3}{rgb}{0.8,0.8,0.8}
\definecolor{v4}{rgb}{0.9,0.5,0.0}
\definecolor{v5}{rgb}{0.8,0.0,0.0}

\newcommand{\za}{\cellcolor{v1}}
\newcommand{\zb}{\cellcolor{v2}}
\newcommand{\zc}{\cellcolor{v3}}
\newcommand{\zd}{\cellcolor{v4}}
\newcommand{\ze}{\cellcolor{v5}}

\makeatother

\begin{document}
\sloppy
\begin{frontmatter}

\title{ABiMed: An intelligent and visual clinical decision support system\\
for medication reviews and polypharmacy management}

\author[label1]{Abdelmalek Mouazer}
\ead{malikmouazer@gmail.com}

\author[label1,label2]{Romain Léguillon}
\ead{romain.leguillon@chu-rouen.fr}

\author[label1]{Nada Boudegzdame}
\ead{nadaboudegzdame@gmail.com}

\author[label3]{Thibaud Levrard}
\ead{tlevrard@eig.fr}
\author[label3]{Yoann Le Bars}
\ead{ylebars@eig.fr}
\author[label3]{Christian Simon}
\ead{csimon@eig.fr}

\author[label1]{Brigitte Séroussi}
\ead{brigitte.seroussi@aphp.fr}

\author[label1,label2]{Julien Grosjean}
\ead{julien.grosjean@chu-rouen.fr}
\author[label1,label2]{Romain Lelong}
\ead{romain.lelong@chu-rouen.fr}
\author[label1,label2]{Catherine Letord}
\ead{catherine.letord@chu-rouen.fr}
\author[label1,label2]{Stéfan Darmoni}
\ead{stefan.darmoni@chu-rouen.fr}

\author[label1,label4]{Matthieu Schuers}
\ead{matthieu.schuers@gmail.com}

\author[label1]{Karima Sedki}
\ead{sedkikarima@yahoo.fr}

\author[label5]{Sophie Dubois}
\ead{sophie.dubois@polesante13.fr}

\author[label5]{Hector Falcoff}
\ead{hector.falcoff@sfr.fr}

\author[label7,label8]{Rosy Tsopra}
\ead{rosytsopra@gmail.com}

\author[label1]{Jean-Baptiste Lamy\corref{cor1}}
\ead{jean-baptiste.lamy@inserm.fr}
\cortext[cor1]{Corresponding author}

\address[label1]{INSERM, Sorbonne Université, Université Sorbonne Paris Nord, Laboratory of Medical Informatics and Knowledge Engineering in e-Health, LIMICS, Paris, France}
\address[label2]{Department of Biomedical Informatics, Rouen University Hospital, France}
\address[label3]{EIG SAS, 92400 Courbevoie, France}
\address[label4]{Département de Médecine Générale, Université de Rouen, France}
\address[label5]{SFTG Recherche (Société de Formation Thérapeutique du Généraliste), Paris, France}
%\address[label6]{}
\address[label7]{Université Paris Cité, Sorbonne Université, Inserm, Centre de Recherche des Cordeliers, F-75006 Paris}
\address[label8]{Department of Medical Informatics, AP-HP, Hôpital Européen Georges-Pompidou, F-75015 Paris, France}
\begin{abstract}
\noindent \emph{Background:} Polypharmacy, \emph{i.e.} taking five
drugs or more, is both a public health and an economic issue. Medication
reviews are structured interviews of the patient by the community
pharmacist, aiming at optimizing the drug treatment and deprescribing
useless, redundant or dangerous drugs. However, they remain difficult
to perform and time-consuming. Several clinical decision support systems
were developed for helping clinicians to manage polypharmacy. However,
most were limited to the implementation of clinical practice guidelines.
In this work, our objective is to design an innovative clinical decision
support system for medication reviews and polypharmacy management,
named ABiMed.

\noindent \emph{Methods:} ABiMed associates several approaches: guidelines
implementation, but the automatic extraction of patient data from
the GP's electronic health record and its transfer to the pharmacist,
and the visual presentation of contextualized drug knowledge using
visual analytics. We performed an ergonomic assessment and qualitative
evaluations involving pharmacists and GPs during focus groups and
workshops.

\noindent \emph{Results:} We describe the proposed architecture, which
allows a collaborative multi-user usage. We present the various screens
of ABiMed for entering or verifying patient data, for accessing drug
knowledge (posology, adverse effects, interactions), for viewing STOPP/START
rules and for suggesting modification to the treatment. Qualitative
evaluations showed that health professionals were highly interested
by our approach, associating the automatic guidelines execution with
the visual presentation of drug knowledge.

\noindent \emph{Conclusions:} The association of guidelines implementation
with visual presentation of knowledge is a promising approach for
managing polypharmacy. Future works will focus on the improvement
and the evaluation of ABiMed.
\end{abstract}
\end{frontmatter}

\section{Background}

Elderly often receive polypharmacy, \emph{i.e.} five drugs or more.
Evidence shows it is a major problem in many countries, including
Canada \citep{Roux2019}, Sweden \citep{Wastesson2019} and France
\citep{Morin2016}. Polypharmacy is both a public health, economic
and ecologic issue. Each new drug administered in polypharmacy increases
the risk of adverse events by 12-18\% \citep{Calderon-Larranaga2012}.

One solution for reducing polypharmacy is medication review (MR),
``\emph{a structured interview with the patient, carried out by the
pharmacist in collaboration with the general practitioner (GP) with
the aim of optimizing patient care}'' \citep{Qassemi2018}. The pharmacist
assesses the treatment, and writes a synthesis with preconizations
for the GP. MR aims in particular at \emph{deprescribing} duplicate
drugs, no longer indicated drugs (\emph{e.g.} statins over 80 in primary
prevention), and dangerous drugs. Other drugs may see their dose changed,
and drugs may also be added, \emph{e.g.} to control adverse events.
Clinical guidelines are available for MR, such as STOPP/START v2 \citep{O'Mahony2015}.

Evidence shows that MR significantly reduces polypharmacy and saves
money, without lowering the quality of care \citep{Zermansky2001},
and can save 273 € per patient-year \citep{Malet-Larrea2017}. MR
may also have a positive ecological impact, by reducing the consumption
of drugs \citep{Davies2022}. In many countries, health insurances
pay pharmacists for performing MR.

However,\textbf{ }few pharmacists are engaged in MR, because they
lack the appropriate knowledge in geriatrics, they fear the reaction
of GPs, and MR is a tedious task. It requires to collect patient data,
including drug orders but also clinical conditions that are often
available only in the GP's electronic health record (EHR). Pharmacists
have to assess the interactions and adverse effects of 5-20 drugs,
to identify inappropriate or missing drugs and to write the synthesis.
Viewing the properties of 5-20 drugs is particularly tedious because
drug databases have been designed to access the properties of a single
drug at a time.

Clinical decision support systems (CDSSs) have been shown to be efficient
in facilitating clinician work, increasing guidelines adherence, and
improving healthcare \citep{Bright2012}. CDSSs have been proposed
for MR. In a literature review \citep{Mouazer2022_revue}, we highlighted
that most execute the rules found in guidelines, and sometimes automatically
extract patient data from EHR. On the other hand, few CDSSs consist
of the visual presentation of selected drug knowledge, \emph{e.g.}
for presenting the summed adverse effects of a drug order.

CDSSs based on the first, \emph{intelligent}, approach include a knowledge
base and an inference engine \citep{Sutton}. The knowledge base can
be formalized in different ways, \emph{e.g.} if/then rules or ontologies.
The inference engine applies the rules to patient data and generates
recommendations for clinicians. Recommendations can be provided in
various ways, \emph{e.g.} alerts or textual reports. For example,
Medsafer \citep{McDonald2019} is an ontology-based system that goes
beyond mere detection of inappropriate drugs. It offers evidence-based
strategies for deprescribing identified inappropriate drugs. N. A.
Zwietering's CDSS \citep{Zwietering2019} implements the STOPP/START
guidelines as if/then rules and generates alerts.

CDSSs based on the second, \emph{visual}, approach rely on drug databases
containing comprehensive drug information, \emph{e.g.} adverse effects
or interactions. The relevant drug knowledge can be displayed to clinicians,
\emph{e.g.} through graphs. It aims at providing efficient access
to information through visual formats that synthesize complex information.
For example, RXplore \citep{visumed-effetsindesirables-frequences}
focuses on adverse effects and presents the information graphically.
Graphsaw \citep{Shoshi2015} visualizes drug interactions and their
associations with various entities, using a network-like structure.

Finally, a \emph{mixed} approach combines both approaches. Few mixed
approaches have been proposed \citep{Mouazer2022_revue}. For example,
KALIS \citep{Shoshi2017} integrates HTA guidelines and the Priscus
inappropriate medication list, but also databases containing molecular
and pharmacological information. KALIS integrates Graphsaw \citep{Shoshi2015},
offering both graphical representations and textual reports. The PRIMA-EDS
system \citep{Sonnichsen2016,Rieckert2018} combines an inference
engine with visual output to check for inappropriate drugs. It employs
the PHARAO2 decision-support system as an inference engine, based
on the EU(7) inappropriate medication list \citep{Renom-Guiteras2015},
as well as drug-oriented databases. The system presents the main adverse
effects in tabular format, alongside detailed textual reports. It
significantly improves the identification and management of inappropriate
drugs.

In the ABiMed project \citep{Mouazer2021_2}, we aim at designing
and evaluating a\textbf{ }CDSS for helping pharmacist to perform MR
and GPs to reduce polypharmacy. ABiMed aims at going beyond state-of-the-art,
by associating guidelines execution with visual approaches, and by
supporting the communication between the pharmacist and the GP, including
the transfer of patient data from the GP's EHR to the pharmacist.

Most published papers on CDSSs for MR focused on evaluation \citep{Mouazer2022_revue},
rather than describing the system design. On the contrary, the objective
of this paper is to describe the ABiMed CDSS, including data exchange,
ontological rule-based system, and original visual interfaces, and
to focus on qualitative evaluations on the software aiming at testing
how desirable are the functionalities we propose.

\section{\label{sec:Methods}Methods}

\subsection{General principles}

The first general principle is to associate in the same CDSS an intelligent
approach, implementing STOPP/START rules, with a visual approach,
consisting of the visual presentation of contextualized drug knowledge,
adapted to the patient profile and treatment.

The second principle is to provide automatic patient data extraction,
to prevent tedious data entry. Extraction is based on the reimbursement
data of the French health insurance or the GP's EHR, when the EHR
software editor integrated support for ABiMed. An EHR editor, EIG
Santé, is a partner of the project and its EHR, éO, will be used to
demonstrate the feasibility of this approach and test whether it is
accepted by GPs (who may be reluctant to share patient data with pharmacists).

The third principle is to display knowledge and recommendations on
either a single drug treatment, but also on two drug treatments (\emph{i.e.}
current treatment \emph{vs} post-MR, which we call the \emph{comparative
mode}). Indeed, most tools related to drug knowledge work at the drug
level, or at the drug order level (for drug interactions). But the
drug level is not appropriated for MR: when a patient takes 5+ drugs,
it is too long for clinicians to read the 5+ corresponding drug pages.
The drug order level is more convenient. However, when suggesting
modifications to the treatment, it does not permit comparing the before-after
MR treatments. For example, one may replace a drug involved in a serious
interaction by another drug, involved in even more serious interactions. 

The fourth principle is to permit a cooperative use of the CDSS, allowing
the pharmacist and the GP to use it simultaneously and to exchange
about the patient. This may turn MR as a more collaborative task,
increasing the involvement of the GP.

\subsection{Intelligent methods}

\subsubsection{Ontologies for structuring patient data}

We previously translated the Observational Medical Outcomes Partnership
- Common Data Model (OMOP-CDM) used to structure EHRs, into an OWL
ontology \citep{Lamy2021_2}. It serves as the basis of the patient
model in ABiMed, and facilitates the management of hierarchical relations
in medical terminologies. The following terminologies were associated:
ICD10 (International Classification of Disease, release 10), ATC (Anatomical
Therapeutical Chemical classification of drugs), LOINC (Logical Observation
Identifiers Names \& Codes), and MedDRA (Medical Dictionary for Regulatory
Activities).

Then, the ontology was enriched for polypharmacy management. Patient
data was divided in 6 categories (OMOP-CDM providing the first three
ones): (1) current drug treatment, including posologies and indications,
(2) clinical conditions of the patient, (3) laboratory tests and exam
results, (4) treatment-related problems identified during the patient
interview (\emph{e.g.} a poor observance), (5) preconizations issued
at the end of the MR (\emph{e.g.} deprescription of a drug), and (6)
chat messages exchanged by the clinicians.

\subsubsection{Standards for exchanging patient data}

We worked with the patient EHR system éO\footnote{\href{https://www.eig.fr/medecin/}{https://www.eig.fr/medecin/} Accessed
12 December 2023}. However, our aim is to be compatible with existing data flows and
to encourage software publishers to endorse our approach. Thus, we
used existing, standard and widely spread file formats for exchanging
patient data. As the project takes place in France, we followed recommendations
from French agencies, especially the CI-SIS specifications\footnote{\href{https://esante.gouv.fr/produits-services/ci-sis}{https://esante.gouv.fr/produits-services/ci-sis}
Accessed 12 December 2023}. The standard we selected may not be the most recent ones, but are
the most used today in France. Following these guidelines, ABiMed
API uses JSON (ISO 21778) as an interchange format, as it is an open
standard file format widely used for data interchange. Numerous tools
are available to manage this file format, facilitating the integration
of ABiMed API.

éO provides two categories of data: the data present in the EHR itself
(including coded and free-text medical data), and the data éO extracts
from reimbursement files from French social security (consisting in
all drugs reimbursed for the patient, whatever the prescriber is),
available from HRI\footnote{\href{https://www.sesam-vitale.fr/hri}{https://www.sesam-vitale.fr/hri}
Accessed 12 December 2023} (\emph{Historique des Remboursements Intégrés}). It is available
in an XML format. It allows getting complementary information concerning
the drugs taken by the patient, including those not prescribed by
the GP but by other physicians, \emph{e.g.} specialists.

Then, éO exports data for ABiMed using the VSM file format\footnote{\href{https://www.has-sante.fr/upload/docs/application/pdf/2013-11/asip_sante_has_synthese_medicale.pdf}{https://www.has-sante.fr/upload/docs/application/pdf/2013-11/asip\_sante\_has\_synthese\_medicale.pdf}
Accessed 12 December 2023} (\emph{Volet de Synthèse Médicale}), based on the HL7 CDA R2 file
format\footnote{\href{http://www.hl7.org/implement/standards/product_brief.cfm?product_id=492}{http://www.hl7.org/implement/standards/product\_brief.cfm?product\_id=492}
Accessed 12 December 2023}. However, we had to make a few modifications to this format. We anonymised
data by removing information such as patient and GP names, \emph{etc}.

\subsubsection{Natural language processing for extracting patient data from free
text}

Automatic extraction of patient data from text is performed using
the Multi-Terminological Concept Extractor (MTCE) \citep{Soualmia,Pereira}.
This semantic annotator has been developed by the Department of Digital
Health from the University Hospital of Rouen (France). It enables
the annotation of texts using terminological and/or ontological concepts
from the Healthcare Ontology and Terminology Portal (HeTOP) \citep{gjhetop}.
A large number of NLP tasks are involved (phrase and word detection,
normalization, \emph{etc.}).

For ABiMed, the primary purpose is to identify patients' clinical
characteristics and lab test results, as required for executing STOPP/START
v2 rules. New functionalities were added to MTCE: the recognition
of conditional and family history information, and negation support.
In fact, many clinical data in consultation texts appear in negative
form. Specific patterns were also designed to retrieves and extract
measures (mostly numerical), \emph{e.g.} lab test results. The aim
was to enable MTCE to produce annotations in the form of $(concept,value)$
pairs, where $concept$ is a LOINC code and $value$ is the numerical
value associated\emph{, e.g.} $(8462\textrm{-}4,\ 95\ mmHg)$, $8462\textrm{-}4$
being the LOINC code for \textit{diastolic blood pressure}.

Finally, MTCE's internal algorithms were updated to improve recall.
In ABiMed, MTCE is used with terminologies that are poorly adapted
to information retrieval due to their complex labels, which are unlikely
to appear in the texts (\emph{e.g.} ``Essential (primary) hypertension''
in ICD10). To overcome this difficulty, it was made possible to exploit
HeTOP's inter-terminological semantic network by internally using
more generic terminologies such as the controlled vocabulary thesaurus
Medical Subject Headings (MeSH). The underlying idea is that MTCE
matches the query not only with the terminologies required in the
ABiMed project, but also to MeSH concepts. The MeSH concepts obtained
are then transcoded into the required terminologies using exact match
relations from HeTOP's semantic network.

\subsubsection{Rule-based system for executing STOPP/START}

The integration of the STOPP/START v2 rules \citep{O'Mahony2015}
involved the formalization and the validation of the rules through
expert consensus. For more details, please refer to \citep{Mouazer2022}.

First, the STOPP/START v2 guidelines were analyzed. They include 114
rules in narrative text format. We identified the necessary logical,
clinical, and attribute elements for detecting potentially inappropriate
medications (PIMs). STOPP rules determine the potential inappropriateness
of a prescription based on the presence or absence of specific clinical
and therapeutic elements, while START rules indicate when a recommended
prescription is absent from the current drug order and needs to be
added. At the end of this step, we developed a formal rule model that
supports all the logical, clinical, and attribute elements we identified.
The rule model relies on the OMOP-CDM-based ontology model described
previously. The general rule format is:

$if\,\,E_{1}\wedge E_{2}\wedge...\wedge\left(U_{1}\vee U_{2}\vee...\right)\wedge\left(...\right)\wedge\neg N_{1}\wedge\neg N_{2}\wedge...$

$then\,\,(stop\,\,or\,\,start)\,\,prescription\,\,P$

\noindent where $E_{i}$, $U_{i}$ and $N_{i}$ are elements (\emph{i.e.}
clinical conditions, drug prescriptions or lab test results) and P
is a drug prescription.

Second, the STOPP/START v2 rules were formalized using that model.
For each rule, this was carried out in three sub-steps: (a) The declaration
of the clinical elements necessary for expressing the rule: prescriptions,
clinical conditions, and lab test results. Each element is associated
with one or more codes in the corresponding terminology (ATC, ICD10
or LOINC, respectively) and can be completed by a set of attributes
(\emph{e.g.} indication or dose, for prescriptions), based on Huibers
\emph{et al.} \citep{Huibers2019}. (b) The writing of the rule logic,
using the above rule format. (c) The writing of the rule alert text.
French translation was based on Lang \emph{et al.} \citep{Lang2015}.
Additionally, comments were added to a rule when the execution of
the rule cannot be fully automatized.

Third, the formalized rules were validated through expert reviews.
Various expert profiles were involved in the review: GP (HF), pharmacist
(SD, RL), researcher in medical informatics (AM, JBL).

Fourth, the structured rules were automatically translated into SPARQL
queries by a Python program. Queries were then executed by the SPARQL
engine in Owlready \citep{Lamy2017_5}.

All 114 STOPP/START v2 rules were considered, with the exception of
the first three STOPP rules (A1, A2, and A3), which are too general
and lack of specificity in terms of drugs.

\subsection{Visual methods}

\subsubsection{Adaptive questionnaire for facilitating patient data entry}

Automatic patient data extraction from EHR is not always possible
(\emph{e.g.} when the GP refuses), and the extracted data may contain
errors and missing elements. In all these situations, manual patient
data entry remains necessary, allowing the pharmacist to verify the
data and complement it if needed. In ABiMed, we designed a questionnaire
targeting the 73 clinical conditions considered in STOPP/START rules.
However, 73-item questionnaire is tedious to fill. Thus, we developed
an adaptive questionnaire that displays only the items strictly mandatory
for executing STOPP/START rules for the current patient \citep{Lamy2023_2}.

For example, rule STOPP J3 recommends to ``Stop beta-blockers in
diabetes mellitus with frequent hypoglycaemic episodes''. There are
3 conditions; one drug: a loop beta-blocker, and two clinical conditions:
diabetes mellitus and hypoglycaemic episodes, related by a logical
AND operator. In the questionnaire, diabetes and hypoglycaemia are
not shown if the patient does not take a beta-blocker. If he/she does,
only diabetes is shown in the questionnaire. If diabetes is checked,
then hypoglycaemia is shown. Consequently, the questionnaire evolves
with the patient data. We showed that this approach reduces the length
of the questionnaire by about two thirds \citep{Lamy2023_2}.

\subsubsection{\label{subsec:Flower-glyphs-and}Flower glyphs and bar charts for
presenting adverse effects}

Adverse effect were described by: (1) nature (a Preferred Term from
MedDRA), (2) frequency (as extracted from the SPCs, on a 5-level scale,
very rare: 0.001-0.01\%, rare: 0.01-0.1\%, uncommon: 0.1-1\%, frequent:
1-10\%, very frequent: >10\%), (3) seriousness (boolean, based on
a list of serious MedDRA terms published by EMA, European Medical
Agency\footnote{\href{https://www.ema.europa.eu/en/documents/other/meddra-important-medical-event-terms-list-version-260_en.xlsx}{https://www.ema.europa.eu/en/documents/other/meddra-important-medical-event-terms-list-version-260\_en.xlsx}
Accessed 12 December 2023}), (4) importance for the elderly (boolean, based on a list published
by the French academy of medicine\footnote{\href{https://www.academie-medecine.fr/effets-indesirables-des-medicaments-chez-les-sujets-ages/}{https://www.academie-medecine.fr/effets-indesirables-des-medicaments-chez-les-sujets-ages/}
Accessed 12 December 2023}).

\begin{figure}
\noindent \begin{centering}
\includegraphics[width=1\columnwidth]{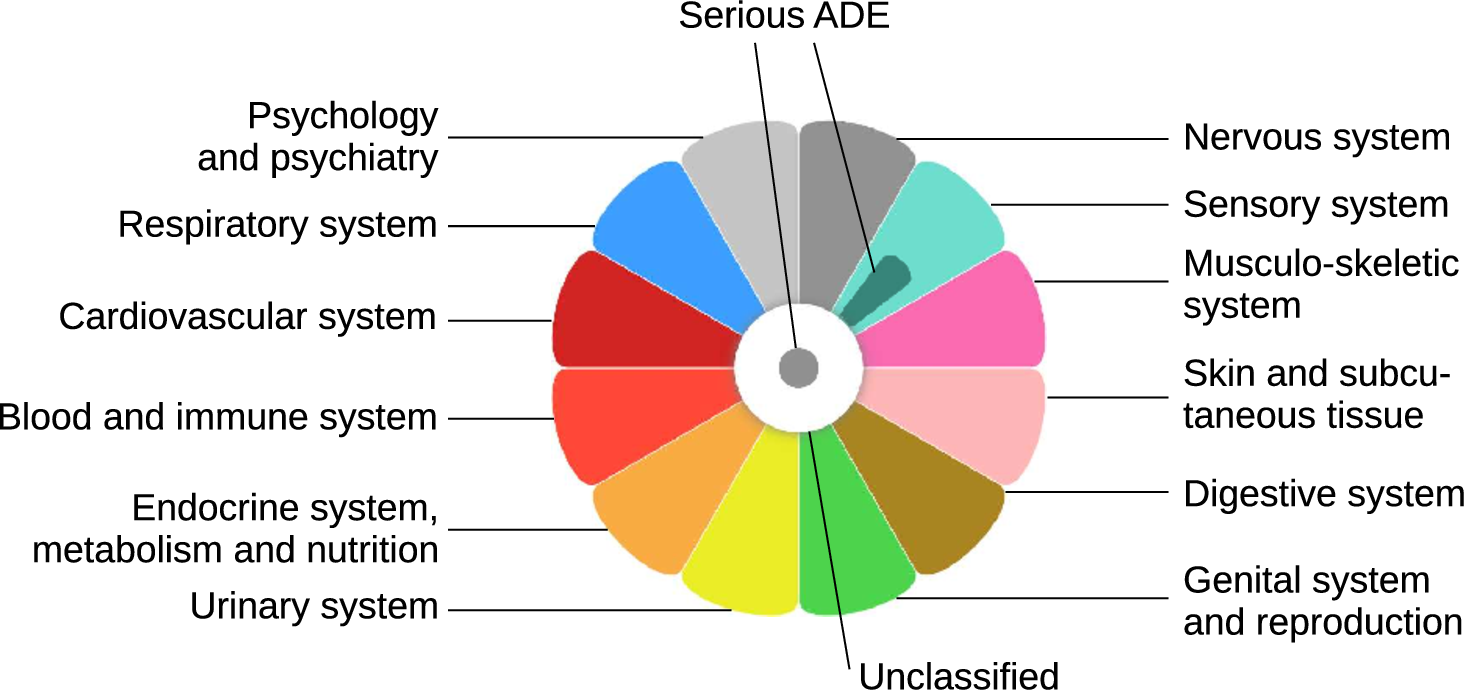}
\par\end{centering}
\caption{\label{fig:fleur}The flower glyph, showing the 13 adverse effect
categories.}
\end{figure}

We designed two types of views for adverse effects. The first is an
overview aggregating adverse effects in general anatomical categories.
In previous works \citep{Lamy2021}, we designed flower glyphs for
the visualization of adverse effect profiles extracted from clinical
trial results. Flower glyphs are similar to bar charts, but bars are
displayed circularly like the petals of a flower. Our glyph has 12
petals, corresponding to 12 general anatomical categories, plus a
central region for a 13\textsuperscript{th} category, unclassified
effects (\emph{e.g.} fatigue). Each petal and region has an inner,
darker, region proportional to the frequency of \emph{serious} adverse
effects. Figure \ref{fig:fleur} shows a flower glyph example and
describes the categories. Each category is associated with a specific
color and direction, chosen to facilitate memorization. We adapted
these flower glyphs to the visualization of the adverse effects described
in the SPCs, for either a single drug or all drugs in a treatment.
For each category, we summed up the frequencies of each drug.

The second view consists of horizontal bar charts. Bars use the same
colors as flower glyphs. Various bar chart views are proposed (see
results section).

\subsubsection{Radial graph visualization for presenting drug interactions}

Drug-drug interactions can be modeled as an undirected labeled graph,
each drug being a node and each drug-drug interaction being an edge
between two nodes. Drug-disease interactions can be simply modeled
as a label on the drug's node. Many methods have been proposed for
graph visualization \citep{Kaufmann2001}. We considered a radial
graph disposition, in which the nodes, representing drugs, are organized
on a circle \citep{Mouazer2020}. Then, node and edge colors are used
to represent drug-disease and drug-drug interactions and their associated
level of gravity.

\subsection{Implementation}

We implemented the CDSS as a web application, in Python (for server)
and Brython (a Javascript-compiled version of Python, for client).
We used WebSockets for client-server communication, permitting the
server to alert the client when patient data have been modified by
another user. This allows several clinicians to collaboratively use
the CDSS for the same patient at the same time.

\subsection{Qualitative evaluations}

\subsubsection{Ergonomic assessment}

The interface of ABiMed was the subject of an ergonomic assessment,
independently by two researchers (RT and JBL). We used two sets of
criteria: the original set by JMC Bastien and DL Scapin \citep{Bastien1993},
and those proposed by P Luzzardi \emph{et al.} \citep{Luzzardi2004}
for information visualization techniques. Problem severity was ranked
on a five-value scale (very minor, minor, average, major, very major).

\subsubsection{Focus groups on prototype}

During two focus groups sessions, mixing GPs and pharmacists, we collected
feedback on a first prototype of ABiMed. A session was organized in
a rural area and the other in an urban area. During these two sessions,
the initial ABiMed prototype was presented and a clinical case was
analyzed collectively using ABiMed. The prototype and its interfaces
were presented, and participants' opinions were collected.

\subsubsection{Workshop with GPs}

We organized a workshop during the French Congress of General Medicine.
Participants were mostly GPs. ABiMed was presented during the workshop,
and then the participants were divided in small groups and asked to
use ABiMed themselves, for analyzing a clinical case. Finally, they
were asked to complete a qualitative evaluation questionnaire. It
included questions about the overall motivation for using ABiMed in
their daily practice, on a scale ranging from 1 to 10, and the opinion
on the usefulness and the presentation of the 4 main tabs of ABiMed,
each expressed on a 5-level qualitative scale.

\section{\label{sec:Results}Results}

\subsection{Architecture}

\begin{figure}[t]
\begin{centering}
\includegraphics[width=1\columnwidth]{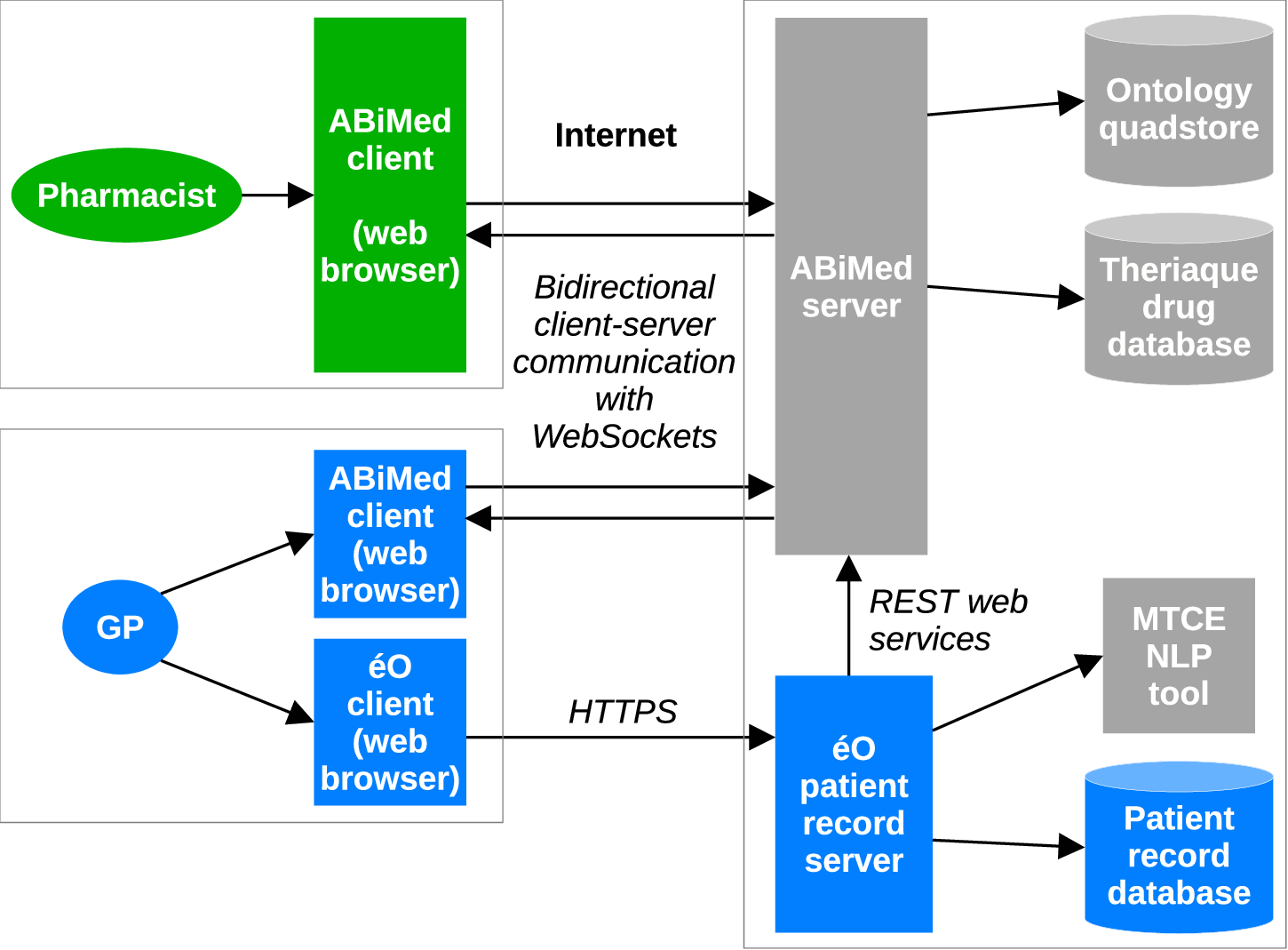}
\par\end{centering}
\caption{\label{fig:Archi}Architecture of the proposed CDSS.}
\end{figure}

Figure \ref{fig:Archi} shows the architecture of ABiMed. It is a
client-server application. The ABiMed server is in relation with web
browser clients, a drug database (Thériaque), with an ontology quadstore
that stores the data and execute STOPP/START rules, and with the éO
EHR server. The éO server is connected to MTCE.

\subsection{CDSS interface}

The CDSS interface includes 8 thematic tabs. A checkbox at the top
of the screen allows switching between the display of drugs as trademarks
or as International Normalized Name (INN). An interactive tutorial
is also proposed. The interface uses colors, but a color-blind friendly
version is available, which uses shades of grays. See Supplementary
file \#1 for additional screenshot.

\subsubsection{Patient data}

This tab displays the patient data. It contains three lists: the list
of prescribed drugs, the list of clinical conditions, and the list
of lab tests and exam results. For each item, the list indicates its
source: manual entry by the pharmacist or the GP, or automatic extraction
from EHR, reimbursement files or textual report. Buttons are proposed
for adding, modifying or removing items. They permit entering patient
data from scratch, or correcting possible errors.

In the drug list, the indications are automatically identified, by
relating the drugs to the clinical conditions, according to the indications
in the Theriaque drug database. The clinician may correct indications.
When there is no indication for a drug, a red label ``Indication???''
is shown, alerting on a potential drug without indication.

\subsubsection{Interview questionnaire}

This tab displays an interview questionnaire that should be filled
by the clinician with the patient. The first part lists the problems
encountered by the patient with his treatment. Five categories of
problem are proposed: (1) suspected adverse drug event, (2) drug intake
difficulty, (3) drug dependency, (4) poor observance, (5) other (free
text).

The second part is focused on the patient lifestyle. It includes checkboxes
related to car driving and addictions (tobacco, alcohol, \emph{etc.}).

The third part is focused on clinical conditions. It is partly redundant
with the clinical conditions in the previous tab, however, only the
conditions relevant for the execution of STOPP/START rules are displayed,
using checkboxes. The checkboxes are organized in 13 anatomical categories
(the same as those for presenting adverse effects). When a checkbox
is checked, a drop-down combo list appears, allowing to select a more
specific ICD10 term (\emph{e.g.} after checking ``diabetes'', one
may choose ``type 1 diabetes'' or ``type 2 diabetes''). This questionnaire
is synchronized with the clinical conditions in the first tab. Moreover,
it is adaptive: the checkboxes shown depend on the drugs taken by
the patient and the clinical conditions previously entered.

\subsubsection{Posologies}

This tab displays the posologies of all drugs, in a table. The columns
are the following: (1) the name of the drug, (2) the current posology
(\emph{i.e.} pre-MR), (3) the official posology, as found in the SPCs
and the Theriaque drug database, (4) the posology preconized by the
pharmacist (\emph{i.e.} post-MR); it can be edited directly in the
table and defaults to the current posology, and (5) the computed day
dose for each active principle in the drug.

In order to reduce the text, the official posologies shown in column
(3) are filtered according to drug indication, drug association, patient
age, renal failure (including the stage and/or renal clearance) and
hepatic failure. ABiMed uses patterns to recognize simple posologies,
such as ``1 morning noon and evening'', ``1 tablet every two days''
or ``1 in case of pain max 6 per day''. If the dose is over the
maximum dose, the maximum dose is highlighted in orange to alert the
clinician. Similarly, when the drug should be taken at a specific
moment (\emph{e.g.} evening) and it is not mentioned in the posology,
that part of the official posology is highlighted. Whenever an active
principle is present in more than one drug, its total dose is shown
in the fifth column, in addition to the per-drug dose.

\begin{figure*}[t]
\noindent \begin{centering}
\includegraphics[width=0.9\textwidth]{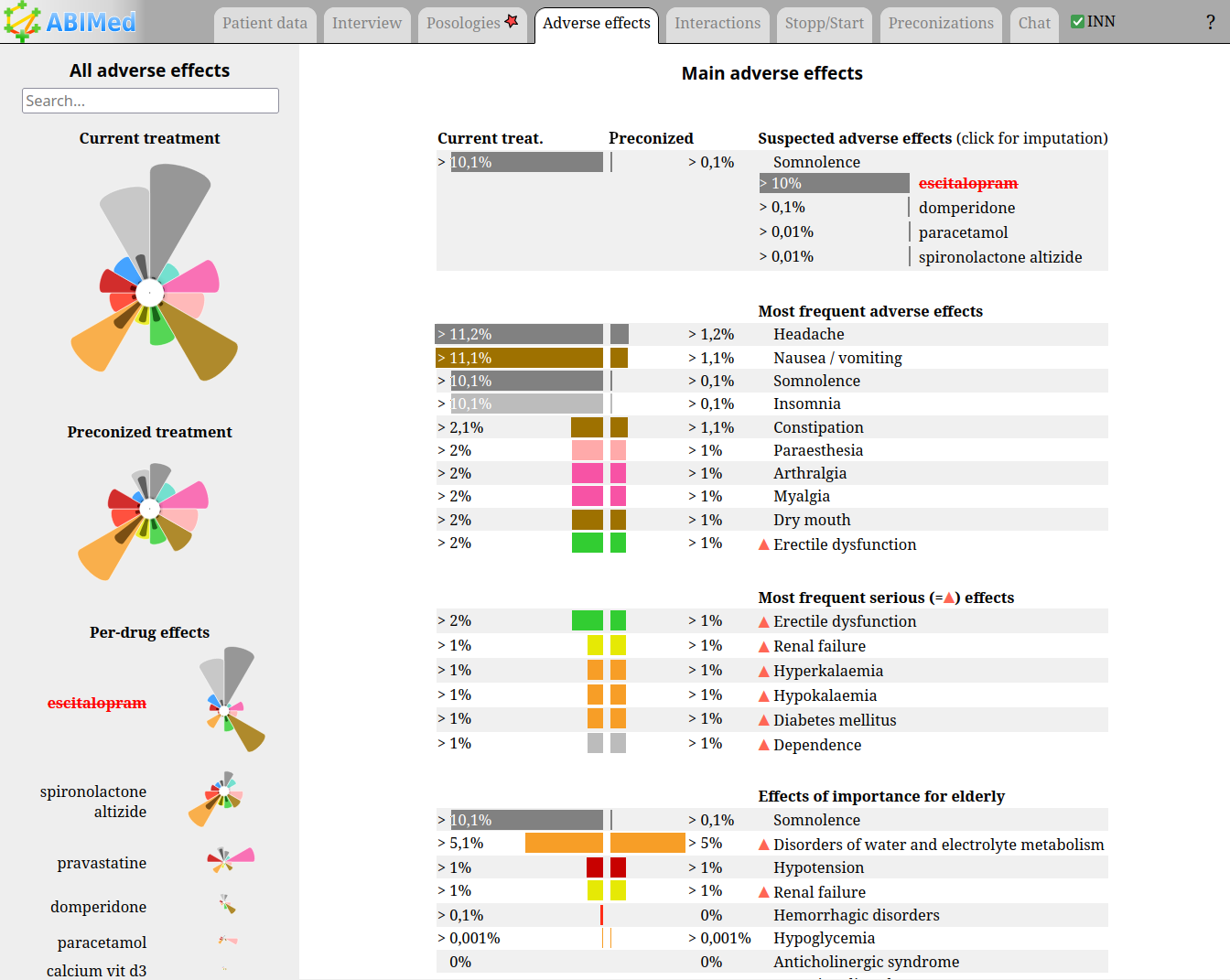}
\par\end{centering}
\caption{\label{fig:ei}Screenshot of the adverse effect tab. The user preconized
to deprescribe escitalopram, thus we are here in comparative mode.}
\end{figure*}

\begin{figure*}[t]
\noindent \begin{centering}
\includegraphics[width=0.9\textwidth]{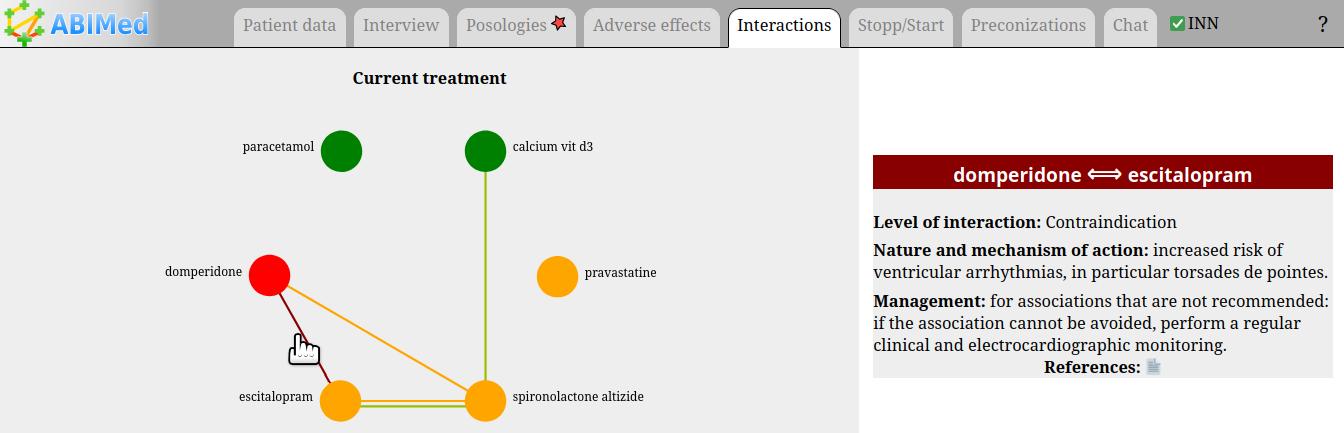}
\par\end{centering}
\caption{\label{fig:interactions}Screenshot of the interaction tab. The user
clicked on the domperidone-escitalopram interaction.}
\end{figure*}

\subsubsection{Adverse effects}

This tab displays the adverse effects of the drug treatment (Figure
\ref{fig:ei}). On the left panel, an overview of the adverse effect
profile of the entire drug treatment is shown, as a flower glyph (see
section \ref{subsec:Flower-glyphs-and}). Smaller, per-drug, flower
glyphs are displayed below, showing the contribution of each drug
to the global profile. Flower glyphs provide, at a glance, an idea
of the general categories of the most frequent adverse effects, for
both serious and non-serious effects. When the mouse is over a petal
or a region, or after a text search, a bubble shows the corresponding
effects in a bar chart series. A triangle is used to mark serious
effects. When an adverse effect is clicked, a per-drug frequency breakdown
is shown.

The right part of the tab displays a summary of the most frequent
and important effects, using four series of bar charts, showing: (1)
the adverse effect suspected in the patient (corresponding to those
entered in the problem list of the previous tab, if any), (2) the
five most frequent adverse effects (considering both serious and non-serious
effects; a higher number of effects may be shown in case of equal
frequency), (3) the five most frequent serious effects, and (4) the
13 effects that are of particular importance for the elderly. All
bar chart series are sorted in decreasing order of frequency.

In comparative mode (\emph{i.e.} when the pharmacist preconized modifications
to the drug treatment), bar chart series display two bars for each
adverse effect, one for the pre-MR treatment and the other for the
post-MR treatment. In addition, a second flower glyph is displayed,
presenting the adverse effect profile of the drug post-MR treatment.
When mouse-hovering a flower glyph, its shape is drawn above the other
glyph, facilitating the identification of small differences. The name
of the added and removed drugs are displayed in blue and in red strikethrough,
respectively.

\begin{figure*}
\begin{centering}
\includegraphics[width=1\textwidth]{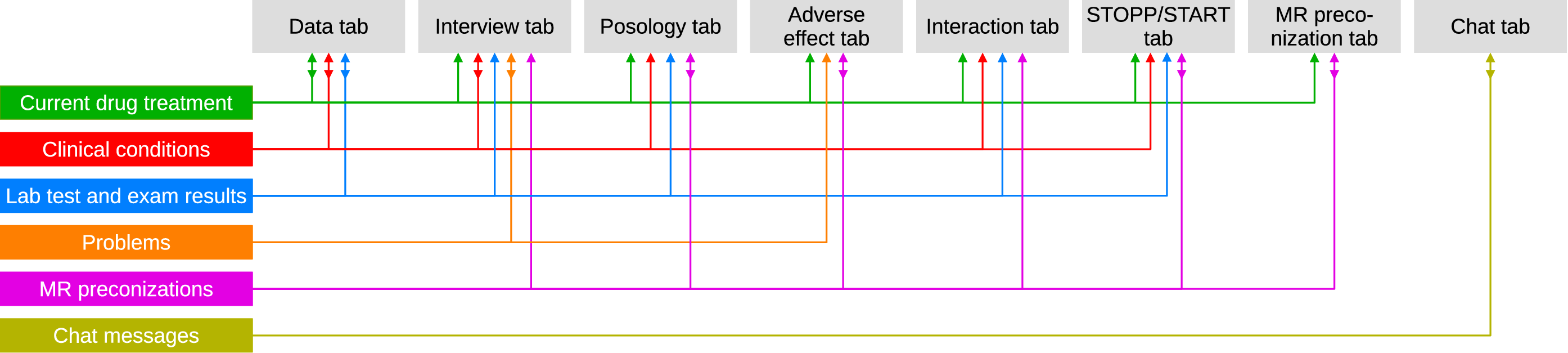}
\par\end{centering}
\caption{\label{fig:Dependencies}Dependencies between the various tabs of
the CDSS and the categories of patient data. One way arrows indicate
that the tab only reads the data. Two-way arrows indicate that it
reads the data and can modify it (for the sake of presentation, the
second arrow is shown next to the first one).}
\end{figure*}

\subsubsection{Interactions}

This tab displays drug-disease and drug-drug interactions using radial
graph visualization (Figure \ref{fig:interactions}). Each drug is
represented by a small colored circle, and all circles are organized
in a large circle. Drug-disease interactions are represented by the
color of the drug circle: red if there is a contraindication, orange
if there is a caution for use (but no contraindication), and green
otherwise. Drug-drug interactions are represented by arcs relating
the two drugs involved; the color of the arcs depends on the severity
of the interaction, with four possible levels. Several arcs are displayed
if there is more than one interaction between two drugs. This visualization
gives an overview of all interactions in the treatment. In particular,
it permits identifying drugs involved in a serious interaction, but
also drugs involved in many interactions of lower seriousness.

By default, the right part of the tab displays the list of the most
important interactions, as text. When the clinician clicks on a drug
circle or an arc, the corresponding detailed information is shown
on the right, including recommendation for taking the interaction
into account and information about the mechanism of action. HTML links
are provided for obtaining full references.

In comparative mode, two interaction circles are shown, one for the
pre-MR treatment and the other for the post-MR treatment. To facilitate
comparison, all drugs are present in both circles (including added
drugs on the first circle, and removed drugs on the second), however,
drugs absent in a treatment are grayed out and their interactions
are not shown.

\subsubsection{STOPP/START rule-based alerts}

This tab shows the STOPP/START v2 rules that match the patient profile.
STOPP rules are shown at the top, ordered by drug. Red/orange/green
traffic signs are used to indicate the three types of rules, respectively:
STOPP rules that are fully automatized, STOPP rules that are not fully
automatized and thus require some intervention of the clinician, and
START rules. In comparative mode, an additional column on the right
shows the rule triggered by the post-MR treatment. Similar drugs before
and after MR are aligned, to facilitate the reading. It allows verifying
that, after substituting a drug by another one, the new drug does
not trigger the same STOPP rule, nor any other ones.

Buttons are proposed for deprescribing drugs matching STOPP rules,
and for prescribing drugs recommended by START rules. These buttons
will update preconizations issued from MR (see next section).

\subsubsection{MR preconizations}

This tab allows the pharmacist to enter the preconizations issued
from MR. Preconizations on the drug treatment can be entered by modifying
the list of prescribed drugs. Six buttons are proposed to (1) signal
a particular problem related with a given drug, or to preconize (2)
the prescription of a new drug, (3) the deprescription of a drug,
(4) the modification of a drug posology, (5) the replacement of a
drug by another, and (6) to cancel a previous preconization.

The drug list displayed in this tab behaves differently than the one
present in the first tab: whenever it is modified, removed drugs are
not removed from the list but displayed in red and strike through,
and added drugs are displayed in blue. Other preconizations, such
as initiating a biological surveillance, can be entered as free text.
Finally, a green button allows validating the MR and sending it to
the GP.

\subsubsection{Chat}

This tab contains a chat, accessible only to the GP and the pharmacist.
It enables an asynchronous text-based communication, specifically
devoted to the management of the current patient.

\subsubsection{Interdependence between tabs}

Figure \ref{fig:Dependencies} shows the dependencies between the
tabs and the patient data categories, illustrating the complexity
and interdependence of ABiMed. Excepted chat, there is no one-to-one
mapping. Many tabs need various patient data categories, sometimes
for very specific items, \emph{e.g.} the posology tab needs lab test
results to find renal clearance.

Whenever patient data is modified, the CDSS automatically updates
the information of all tabs as needed, by extracting the appropriate
drug information and executing STOPP/START rules again. Tabs having
new contents are highlighted with a red star, to alert the clinician,
\emph{e.g.}, prescribing a new drug in the MR preconizations tab will
add a red star to the STOPP/START tab if the prescribed drug triggers
any STOPP rule.

Additionally, the CDSS can be used cooperatively by multiple users.
A pharmacist and a GP can display the same patient. They can exchange
\emph{via} the chat, and both can perform modifications (\emph{e.g.}
correct patient data or modify the post-MR treatment). Other users
will be warned of the changes by the apparition of red stars.

\begin{figure*}[t]
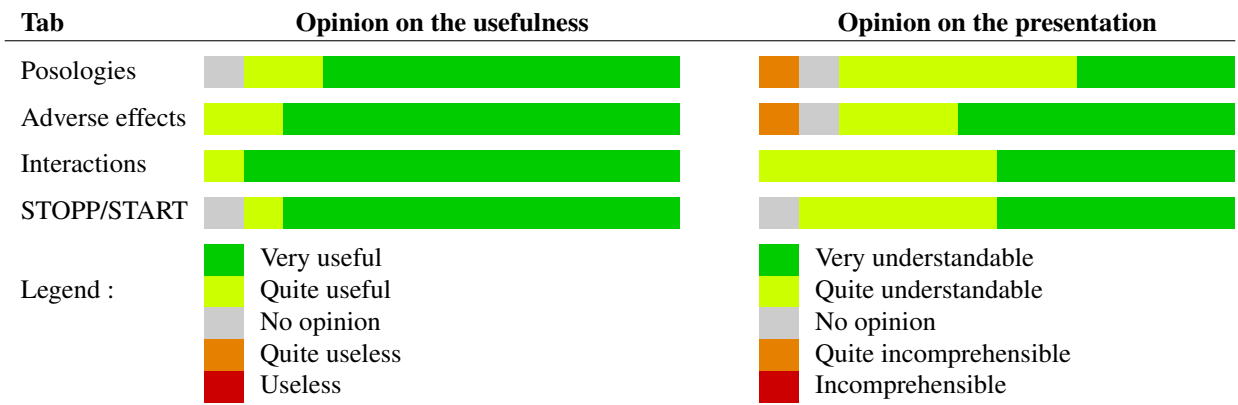

\begin{centering}
\begin{tabular}{l>{\centering}p{0.1cm}>{\centering}p{0.1cm}>{\centering}p{0.1cm}>{\centering}p{0.1cm}>{\centering}p{0.1cm}>{\centering}p{0.1cm}>{\centering}p{0.1cm}>{\centering}p{0.1cm}>{\centering}p{0.1cm}>{\centering}p{0.1cm}>{\centering}p{0.1cm}>{\centering}p{0.1cm}>{\centering}p{0.1cm}>{\centering}p{0.1cm}>{\centering}p{0.1cm}>{\centering}p{0.1cm}>{\centering}p{0.1cm}>{\centering}p{0.1cm}>{\centering}p{0.1cm}>{\centering}p{0.1cm}>{\centering}p{0.1cm}>{\centering}p{0.1cm}>{\centering}p{0.1cm}>{\centering}p{0.1cm}>{\centering}p{0.1cm}>{\centering}p{0.1cm}}
\textbf{Tab} & \multicolumn{12}{c}{\textbf{Opinion on the usefulness}} &  &  & \multicolumn{12}{c}{\textbf{Opinion on the presentation}}\tabularnewline
\hline 
\noalign{\vskip0.2cm}
Posologies & \zc  & \zb  & \zb  & \za  & \za  & \za  & \za  & \za  & \za  & \za  & \za  & \za  &  &  & \zd  & \zc  & \zb  & \zb  & \zb  & \zb  & \zb  & \zb  & \za  & \za  & \za  & \za \tabularnewline
\noalign{\vskip0.2cm}
Adverse effects & \zb  & \zb  & \za  & \za  & \za  & \za  & \za  & \za  & \za  & \za  & \za  & \za  &  &  & \zd  & \zc  & \zb  & \zb  & \zb  & \za  & \za  & \za  & \za  & \za  & \za  & \za \tabularnewline
\noalign{\vskip0.2cm}
Interactions & \zb  & \za  & \za  & \za  & \za  & \za  & \za  & \za  & \za  & \za  & \za  & \za  &  &  & \zb  & \zb  & \zb  & \zb  & \zb  & \zb  & \za  & \za  & \za  & \za  & \za  & \za \tabularnewline
\noalign{\vskip0.2cm}
STOPP/START & \zc  & \zb  & \za  & \za  & \za  & \za  & \za  & \za  & \za  & \za  & \za  & \za  &  &  & \zc  & \zb  & \zb  & \zb  & \zb  & \zb  & \za  & \za  & \za  & \za  & \za  & \za \tabularnewline
\noalign{\vskip0.2cm}
 & \za  & \multicolumn{11}{l}{Very useful} &  &  & \za  & \multicolumn{11}{l}{Very understandable}\tabularnewline
Legend : & \zb  & \multicolumn{11}{l}{Quite useful} &  &  & \zb  & \multicolumn{11}{l}{Quite understandable}\tabularnewline
 & \zc  & \multicolumn{11}{l}{No opinion} &  &  & \zc  & \multicolumn{11}{l}{No opinion}\tabularnewline
 & \zd  & \multicolumn{11}{l}{Quite useless} &  &  & \zd  & \multicolumn{11}{l}{Quite incomprehensible}\tabularnewline
 & \ze  & \multicolumn{11}{l}{Useless} &  &  & \ze  & \multicolumn{11}{l}{Incomprehensible}\tabularnewline
\end{tabular}
\par\end{centering}
\caption{\label{fig:Opinion-tab}Opinion on the main tabs of ABiMed of the
12 GPs participating to the workshop.}
\end{figure*}

\subsection{Qualitative evaluation results}

\subsubsection{Ergonomic assessment}

Fifty ergonomic problems were identified (48 from the criteria of
JMC Bastien and DL Scapin, and 2 from the criteria of P Luzzardi \emph{et
al.}), including 11 very minor, 24 minor, 14 average, 1 major and
0 very major. 30 have been corrected, including 8 very minor, 18 minor,
3 average and 1 major. Many problems were: (1) of very low severity,
\emph{e.g.} the absence of handling of the ``escape'' key to close
dialog boxes, (2) not directly related to the decision support activity,
\emph{e.g.} the absence of a ``password lost'' functionality, or
(3) not problematic for performing evaluations on a limited number
of patients, \emph{e.g.} the impossibility to sort the patient list
by date. Other problems would require important developments, \emph{e.g.}
adding an ``undo'' functionality when modifying the patient data.

\subsubsection{Focus groups on prototype}

The rural focus group involved 4 pharmacists (gender: 2 male, 2 female,
age from 35 to 45 years) and 4 GPs (all male, age 32-66). The urban
focus group involved 4 pharmacists (3 male, 1 female, age 45-56) and
4 GPs (2 male, 2 female, 36-49). The participants were unanimous in
considering that ABiMed would be useful in practice and that the interfaces
were satisfactory. One said: “It’s practical, and visual too!”. The
provision of contextualized drug knowledge from different sources,
as well as the care taken in their visualization, were very appreciated.
One participant said: ``Interesting, in particular, to identify drugs
which do not have major interactions, but which participate in a large
number of interactions''. The participants thought they would use
it in their practice to analyze a prescription or before prescribing
a new medication. A pharmacist internship supervisor intended to use
it as an educational tool, saying: “For everyday practice, it will
be a good tool, even just to check something”. ABiMed was found to
be a good basis for promoting exchanges between doctors and pharmacists,
but also with patients. A point of vigilance concerned the notifications
of messages between the GP and the pharmacist, multiplying messages
being at the risk of losing responsiveness.

\subsubsection{Workshop with GPs}

\begin{figure}[t]
\begin{centering}
\includegraphics[width=0.9\columnwidth]{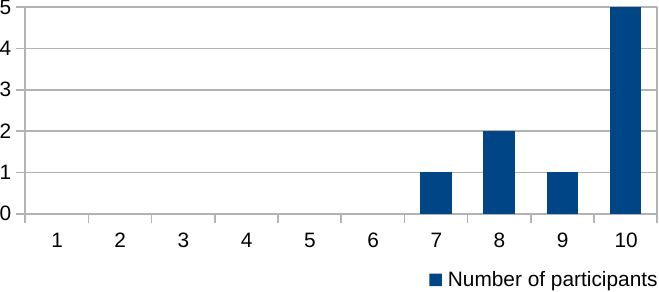}
\par\end{centering}
\caption{\label{fig:Opinion-motiv}Opinion of the GP participating in the workshop
on their motivation to use ABiMed (0: no motivation at all, 10: maximum
motivation).}
\end{figure}

The workshop included 13 participants (only 12 completed the questionnaire).
Regarding gender, 6 were male and 6 female. Seven were practicing
GPs (seniority ranging from 1 to 36 years), one was a retired GP,
and three were working in agencies (Health Insurance and French National
Health Agency). Participants were enthusiastic with regard to the
proposed system. The average score for motivation to use ABiMed in
daily practice was 9.1 on a scale of 1 to 10 (Figure \ref{fig:Opinion-motiv};
three GPs, including those working in agencies, did not reply).

Figure \ref{fig:Opinion-tab} shows the results for the questions
on the 4 main tabs, posology, adverse effects, interactions and STOPP/START
rules. All tabs were judged highly useful, but the opinion on their
presentation is slightly more mixed. Surprisingly, the tabs that use
complex visual analytics (\emph{i.e.} adverse effects and interactions)
were not perceived as less understandable.

\section{\label{sec:Discussion}Discussion}

In this paper, we described the design of ABiMed, an intelligent and
visual decision support system for medication reviews and polypharmacy
management. ABiMed implements the rules from the STOPP/START v2 guidelines,
and proposes contextualized drug knowledge with a visual presentation.
The system also permits a collaborative multi-user usage, similar
to online office suites. ABiMed was evaluated qualitatively by pharmacists
and GPs during two focus groups and a workshop. The results show that
health professionals are interested in the proposed system, and that,
despite the use of complex visual analytics, it remains understandable.

We worked with EIG Santé as industrial partner, which develops the
éO software for physicians. While it might have been more obvious
to work with an editor of pharmacy management software, the amount
of clinical data available in such software is very limited. Therefore,
we chose to work with a vendor of medical practice management software,
allowing the extraction of clinical data from the GP's database. We
currently worked with a single vendor, in order to establish a proof-of-concept,
before considering the integration of ABiMed with additional software
from other vendors. However, convincing all vendors is expected to
be challenging.

The main limits of ABiMed are the lack of quantitative performance
evaluation and clinical use in practice. It also requires a high degree
of collaboration between pharmacists and GPs, which may be difficult
to achieve. Regarding the adverse effects tab, we summed the frequencies
of all adverse effects for each drug, however, in practice, it can
lead to high percentages and the frequencies may not be additive.
Nevertheless, those summed frequencies, although imperfect, gives
an indication of the risk of adverse events.

In the literature \citep{Mouazer2022_revue}, most CDSSs for MR were
focused on the implementation of guidelines. ABiMed also proposes
that, but goes beyond state-of-the-art, with the addition of visual
tools for presenting contextualized drug knowledge. Moreover, in the
literature, CDSSs devoted to community pharmacists were not connected
to EHRs, because the pharmacist has no direct access to GP's EHR.
Consequently, the collaboration we propose between pharmacists and
GPs, based on the transfer of patient data from the GP to the pharmacist,
is innovative. We also proposed a comparative mode, showing both the
analyses of the pre-MR and the post-MR treatment.

The main perspective is the evaluation of ABiMed. We are currently
conducting a performance evaluation with pharmacists on clinical cases,
under controlled conditions, aimed at showing that ABiMed leads to
better MR. We also work on the evaluation of the rule-based system
on retrospective patient data. In a next step, we plan to evaluate
ABiMed in a clinical trial with real patients, associating both pharmacists
and GPs.

In future work, we would like to explore the use of argumentation
\citep{Scheuer2010} as a way to structure communication between the
pharmacist and the GP. Actually, effective communication between the
pharmacist and the GP is important for MR. As both share a common
goal, \emph{i.e.} improving the health of the patient, exchanging
arguments may resolve most of the disagreements and facilitate MR.
In this context, drug interactions, adverse effects, STOPP/START rules,
but also patient preferences, can be considered in the process of
argumentation for justifying pharmaceutical interventions. Arguments
may even be extracted from natural language messages in the chat.
Another perspective is the addition of a ``life line'', \emph{i.e.}
a chronological view showing the patient clinical conditions and drug
prescriptions on a temporal axis \citep{visumed-lignedevie2}. Such
a view might improve the understanding of the patient history. However,
it requires temporal data, which may be difficult to obtain, especially
for the pharmacist (\emph{e.g.} when extracting drug treatment from
health insurance reimbursement, only the drug delivered in the last
months are present, thus the initial prescription date cannot be obtained).
Finally, the implementation of STOPP/START v3 is another planned work.

\section{\label{sec:Conclusion}Conclusions}

In conclusion, we proposed an intelligent and visual clinical decision
support system for medication review and polypharmacy management.
It relies on (1) the automatic extraction of patient data from the
GP's EHR and its transfer to the pharmacist, (2) the implementation
of the STOPP/START rules, and (3) the presentation of contextualized
drug knowledge using visual analytics. Qualitative evaluations showed
that clinicians were highly interested. Future works will focus on
the evaluation of the system and its improvements.

\section*{List of abbreviations}
\begin{lyxlist}{00.00.0000}
\item [{ATC}] anatomical therapeutical chemical classification of drugs
\item [{CDSS}] clinical decision support system
\item [{EHR}] electronic health record
\item [{GP}] general practitioner
\item [{HeTOP}] healthcare ontology and terminology portal
\item [{ICD10}] iInternational classification of disease, release 10
\item [{INN}] international normalized name
\item [{LOINC}] logical observation identifiers names \& codes
\item [{MedDRA}] medical dictionary for regulatory activities
\item [{MeSH}] medical subject headings
\item [{MR}] medication review
\item [{MTCE}] multi-terminological concept extractor
\item [{OMOP-CDM}] observational medical outcomes partnership - common
data model
\item [{PIMs}] potentially inappropriate medications
\end{lyxlist}

\section*{Declarations}

\subsection*{Ethics approval and consent to participate}

The focus groups and workshop studies presented here included no patient
and no real patient data. Thus, according to the French regulation,
no ethics approval is required.

\subsection*{Consent for publication}

Not applicable

\subsection*{Availability of data and materials}

The dataset generated and analyzed during the current study, with
the data collected during the workshop, is available in this published
article's supplementary files.

\subsection*{Competing interests}

Stefan Darmoni and its team sold the MTCE software (which is used
in the present work) via the Alicante company.

\subsection*{Funding}

This work was funded by the French Research Agency (ANR) through the
ABiMed project {[}Grant No. \mbox{ANR-20-CE19-0017}{]}.

\subsection*{Authors' contributions}

AM, R Léguillon, NB, YLB, BS, JG, R Lelong, CL, SD, KS and JBL contributed
to the conceptualization and the methods. AM, TL, CS, R Lelong and
JBL developed the software. AM, R Léguillon, MS, SD, HF, RT and JBL
contributed to the evaluations. R Léguillon, NB, YLB, BS, R Lelong,
CL, KS, SD, HF and JBL wrote the first draft of the manuscript. All
authors read and approved the final manuscript.

\subsection*{Acknowledgements}

Not applicable

\section*{Supplementary files}

Supplementary file \#1: Screenshots of the CDSS.

Supplementary file \#2: data table with data collected through questionnaires
during the workshop.

\bibliographystyle{vancouver}
\bibliography{biblio_ama,biblio_ecmt}

\end{document}